\documentclass[twocolumn,aps,floatfix,showpacs,superscriptaddress]{revtex4}
\newcommand{\be}{\begin{equation}}
\newcommand{\ee}{\end{equation}}
\newcommand{\bea}{\begin{eqnarray}}
\newcommand{\eea}{\end{eqnarray}}

\usepackage{graphicx}
\usepackage{bm}
\begin{document}

\begin{abstract}

\pacs{03.75.Dg, 03.75.Lm, 37.25.+k}

We demonstrate the operation of an atom interferometer based on a
weakly interacting Bose-Einstein condensate. We strongly reduce
the interaction induced decoherence that usually limits
interferometers based on trapped condensates by tuning the s-wave
scattering length almost to zero via a magnetic Feshbach
resonance. We employ a $^{39}$K condensate trapped in an optical
lattice, where Bloch oscillations are forced by gravity. With a
control of the scattering length better that 0.1 $a_0$ we achieve
coherence times of several hundreds of ms. The micrometric sizes
of the atomic sample make our sensor an ideal candidate for
measuring forces with high spatial resolution. Our technique can
be in principle extended to other measurement schemes opening new
possibilities in the field of trapped atom interferometry.
\end{abstract}

\title{Atom interferometry with a weakly-interacting Bose Einstein condensate}
\author{M. Fattori}
\affiliation{LENS and Dipartimento di Fisica, Universit\`a di
Firenze,
  and INFM-CNR\\  Via Nello Carrara 1, 50019 Sesto Fiorentino, Italy }
\affiliation{Museo Storico della Fisica e Centro Studi e Ricerche
'Enrico Fermi' ,Compendio del Viminale, 00184 Roma, Italy }

\author{C. D'Errico}
\affiliation{LENS and Dipartimento di Fisica, Universit\`a di
Firenze,
  and INFM-CNR\\  Via Nello Carrara 1, 50019 Sesto Fiorentino, Italy }
\affiliation{INFN, Sezione di Firenze, Via Sansone 1, 50019 Sesto
Fiorentino, Italy }

  \author{G. Roati}
\affiliation{LENS and Dipartimento di Fisica, Universit\`a di
Firenze,
  and INFM-CNR\\  Via Nello Carrara 1, 50019 Sesto Fiorentino, Italy }
\affiliation{INFN, Sezione di Firenze, Via Sansone 1, 50019 Sesto
Fiorentino, Italy }

  \author{M. Zaccanti}
\affiliation{LENS and Dipartimento di Fisica, Universit\`a di
Firenze,
  and INFM-CNR\\  Via Nello Carrara 1, 50019 Sesto Fiorentino, Italy }

  \author{M. Jona-Lasinio}
\affiliation{LENS and Dipartimento di Fisica, Universit\`a di
Firenze,
  and INFM-CNR\\  Via Nello Carrara 1, 50019 Sesto Fiorentino, Italy }

  \author{M. Modugno}
\affiliation{LENS and Dipartimento di Fisica, Universit\`a di
Firenze,
  and INFM-CNR\\  Via Nello Carrara 1, 50019 Sesto Fiorentino, Italy }
  \affiliation{INFN, Sezione di Firenze, Via Sansone 1, 50019 Sesto
Fiorentino, Italy }
  \affiliation{Dipartimento di Matematica Applicata, Universit\`a di Firenze, Italy}
  \affiliation{BEC-INFM Center, Universit\`a di Trento, I-38050 Povo, Italy}

\author{M. Inguscio}
\affiliation{LENS and Dipartimento di Fisica, Universit\`a di
Firenze,
  and INFM-CNR\\  Via Nello Carrara 1, 50019 Sesto Fiorentino, Italy }
\affiliation{INFN, Sezione di Firenze, Via Sansone 1, 50019 Sesto
Fiorentino, Italy }

  \author{G. Modugno}
\affiliation{LENS and Dipartimento di Fisica, Universit\`a di
Firenze,
  and INFM-CNR\\  Via Nello Carrara 1, 50019 Sesto Fiorentino, Italy }
\affiliation{INFN, Sezione di Firenze, Via Sansone 1, 50019 Sesto
Fiorentino, Italy }

\maketitle

A Bose Einstein condensate (BEC) is the ideal candidate for atom
interferometry because it offers the ultimate control over phase
and amplitude of a matter wave. Its macroscopic coherence length
\cite{KInterf} allows high phase contrast in interferometric
experiments. Its confinement in atomic traps or waveguides can
lead to the realization of atom interferometers with high spatial
resolution \cite{interferometry, Shin}. Unfortunately in high
density trapped condensed clouds interaction induces phase
diffusion \cite{CastinDalib}, thus seriously limiting the
performances of a BEC atom interferometer. In order to avoid the
deleterious effect of interaction, atom interferometers for high
precision measurement use free falling dilute samples of non
degenerate atoms \cite{Peters, Giroscopio}. The main drawbacks are
the limited interrogation time (0.5~s) due to the finite size of
the apparatus and the poor spatial resolution of this type of
sensors. Using fermionic atoms instead of bosons represents one
possibility to have access to trapped interferometry with a
degenerate gas \cite{Roati}. In fact due to Pauli exclusion
principle the atomic scattering cross section at sufficiently low
temperatures is fully suppressed. However the quantum pressure
limits the spatial resolution, and the momentum spread reduces the
interference contrast. Another way to reduce the effect of the
interaction is to realize a number squeezed splitting
\cite{KSqueezing, KasevichSqueezing}. In this way coherence times
are increased at the expense of the interference signal
visibility. Despite the fundamental limit represented by
interaction induced decoherence, several groups are performing
experiments with trapped BECs \cite{KSqueezing, KasevichSqueezing,
Zimmermann, Schumm, Sackett}, in the challenging search for the
``ideal" interferometer.

In this Letter we demonstrate the conceptually simplest solution
to the long standing problem of interaction induced decoherence in
BEC interferometers. We show how, by properly tuning the
interaction strength in a quantum degenerate gas of $^{39}$K
\cite{k39} by means of a broad magnetic Feshbach resonance
\cite{FSK39}, we can greatly increase the coherence time of an
atom interferometer. By achieving almost vanishing values of the
s-wave scattering length, we demonstrate trapped atom
interferometry with a weakly interacting BEC.

The interferometer we adopted is based on a multiple well scheme
\cite{interferometry, Roati, biraben}. The condensate is
adiabatically loaded in a sinusoidal potential with period
$\lambda /2$, realized with an optical standing wave of wavelength
$\lambda$. In the presence of an external force $F$, the
macroscopic wavefunction $\psi$ of the condensate can be described
as a coherent superposition of Wannier Stark states $\phi_i$
\cite{Meystre}, parametrized with the lattice site index $i$,
characterized by complex amplitudes of module $\sqrt{\rho_i}$ and
phase $\theta_i$, $\psi = \sum_i \sqrt{\rho_i} exp(j \theta_i)
\phi_i$ \cite{Korsch}. In the absence of interaction the phase of
each state evolves according to the energy shift induced by the
external potential, i.e. $\theta_i = F\lambda i t/2$. By releasing
the cloud from the lattice, the macroscopic interference between
different Wannier Stark states gives rise to the well known Bloch
oscillations of the density pattern, with period $t_{bloch} = 2h/F
\lambda$. A measurement of the frequency of such oscillations
allows a direct measurement of the external force. Generally
interactions give rise to a complex system of non linear equations
for $\rho_i$ and $\theta_i$. In the weakly interacting limit the
$\rho_i$ don't change and, in addition to $\theta_i$, extra phase
terms $\theta_i'$, proportional to the local interaction energy,
are accumulated, i.e. $\theta_i' \propto g \rho_i t/h$, where $g$
is the interaction strength proportional to the scattering length
$a_s$. This causes phase diffusion and destruction of the
interference pattern. In this Letter we demonstrate that by tuning
$a_s$ almost to zero it is possible to reduce the values of
$\theta_i'$ and extend the coherence time of the interferometer.

\begin{figure}
\includegraphics[width=\columnwidth]{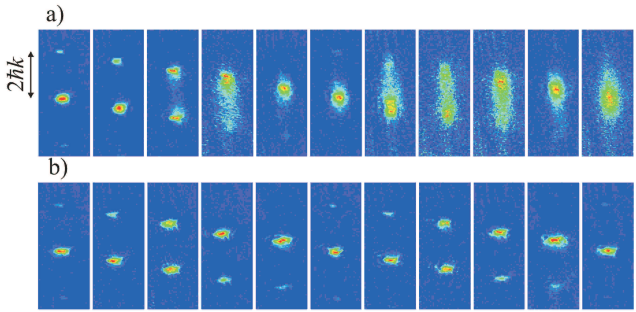}
\caption{(Color online) Bloch oscillations from 0 to 4~ms, in
steps of 0.4~ms, for a condensate with a) 100~$a_0$ and b) 1~$a_0$
scattering length. The picture shows absorption images of the
cloud after release from the lattice. The expansion lasts 12.5~ms
and the scattering length is changed to -33~$a_0$ only 3~ms before
image acquisition. The arrow shows the momentum separation $2
\hbar k$ between the interference peaks, where $k = 2\pi
/\lambda$, with $\lambda$ the lattice wavelength.} \label{fig1}
\end{figure}

Our experimental apparatus has been described in detail elsewhere
\cite{k39}. Using sympathetic cooling with $^{87}$Rb and other
techniques similar to the ones developed for other potassium
isotopes \cite{Roati2002, science, deSarlo, k39}, we produce a
$^{39}$K BEC with $4\times 10^4$ atoms in the absolute ground
state $|F=1, m_F=1 \rangle$. The collisional properties of
$^{39}$K have been extensively studied in \cite{FSK39}. Due to the
negative background scattering length of $^{39}$K
($a_{bg}=-33~a_0$), the final stage of evaporation is performed in
a crossed optical dipole trap at a magnetic field of 395 G. At
this field a broad Feshbach resonance ($\Delta=-52$~G, center of
the resonance $B_c=402.4$~G) allows the homo-nuclear scattering
length of K to be large and positive ($\approx 200~a_0$) and a
stable condensate can form. Taking into account the dependence of
$a_{s}$ on the magnetic field $a_s(B)=a_{bg}(1-(\Delta/(B-B_c)))$,
the possibility of tuning with high accuracy the interaction from
repulsive to attractive is evident. In particular, around
$B_{zc}=B_c+\Delta$, i.e. the point where $a_s$ vanishes,
$a_s(B)\sim a_{bg}/\Delta \cdot (B-B_{zc})$ and a fine control of
the scattering length ($0.6~a_0/$~G) is possible. Our magnetic
field stability, better than 100~mG, allows a reduction of $a_s$
by nearly a factor of $10^3$, down to the 0.06 $a_0$ level. This
high degree of tunability is possible only in few other atomic
species where the ratio $a_{bg}/\Delta$ is favourably small
\cite{Lithium, cesium2}.

The condensate is produced in a horizontal crossed optical dipole
trap with frequencies $(\nu_x, \nu_y, \nu_z)=(85, 105, 100)$~Hz.
After production of a degenerate sample we adiabatically tune
$a_s$ to the desired value and we load the atoms in a vertical
optical lattice with depth $s E_R$, where $s \sim 6$ and
$E_R=h^2/2m\lambda^2$ is the recoil energy from absorption of a
lattice photon ($\lambda = 1032$~nm). In order to avoid an abrupt
change of the horizontal confinement from the combined potential,
i.e. the trap plus the lattice, to the lattice only, we apply a
levitating magnetic field gradient and decrease adiabatically the
trapping confinement to $(44, 76, 43)$~Hz. When the crossed dipole
trap and the levitating field are switched off, Bloch oscillations
start in the vertical lattice with radial confinement
$\nu_x=\nu_y=44$~Hz.

A first demonstration of our capability to extend the coherence
times of the interferometer by tuning $a_s$ is presented in Fig.
1. Here we report the absorption images of BECs released from the
lattice, after different times of oscillations performed at
100~$a_0$ and 1~$a_0$ scattering length. The $a_s = 100~a_0$
measurement shows the performances of the interferometer with a
typical interacting condensate \cite{Kasevich2}. After two Bloch
periods the interference pattern is drastically broadened. Instead
the measurement with $a_s=1~a_0$ shows no discernible broadening
on this time scale, displaying how the interaction induced
decoherence can be strongly suppressed with our method.

To analyze quantitatively the effect of the tuning of the
interaction we repeat the same experimental sequence for different
$a_s$, measuring the vertical width of the central peak at integer
times of the Bloch period (Fig. 2). This study is limited to a
range of values of $a_s$ ($29~a_0 - 1.3~a_0$) where interaction is
the main source of decoherence (see below). Notice that initially
the widths increase linearly with time. This is a direct
consequence of the phase terms $\theta_i' \propto g \rho_i t/h$,
with $\rho_i =$const, that evolve linearly in time \cite{Korsch}.
Later on, when the momentum distribution of the condensate
occupies the whole first Brillouin zone, the widths saturate. From
a numerical calculation of $\rho_i$ and the theoretical analysis
described in \cite{Korsch}, we derive the interference pattern
between the Wannier Stark states in the momentum space and
determine the width of the central peak as a function of time. In
Fig. 3 we compare the theoretical decoherence rates with the
measured ones as a function of $a_s$. For this range of values an
almost linear behavior is found. The possibility to preserve the
visibility of the interference pattern for hundreds of ms when
$a_s$ is tuned down to $\sim 1~a_0$ is astonishing. Notice that
our theoretical analysis doesn't take into account the expansion
of the cloud in the presence of interaction when the atoms are
released from the lattice. In addition it completely neglects
energetic and dynamical instability effects \cite{deSarlo2}. These
are good approximations for small values of $a_s$.

\begin{figure}
\includegraphics[width= \columnwidth,clip]{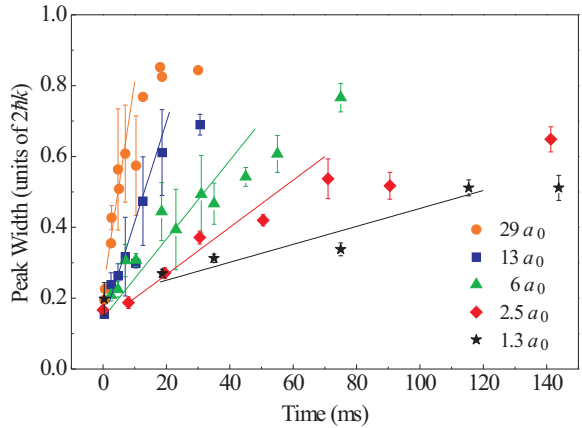}
\caption{(Color online) Vertical $1/\sqrt{e}$ width of the central
peak of the density profile as a function of the Bloch oscillation
time, at integer multiples of the Bloch period after 12.5 ms of
expansion from the trap. The measurement is performed for
different values of the scattering length, listed on the legend,
with $4 \cdot 10^4$ atoms on average. Only at the end of the
expansion the external magnetic field is switched off. The lines
are a fit to the data excluding the point at t=0 ms, where imaging
saturation effects can occur, up to times when the peak width is
approximately $0.5\times 2\hbar k$. } \label{fig1}
\end{figure}

\begin{figure}
\includegraphics[width=\columnwidth
,clip]{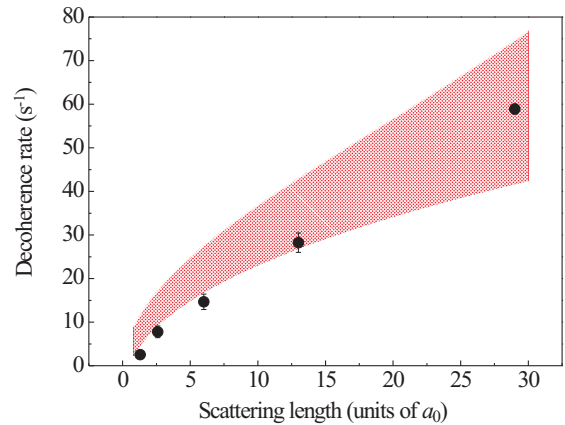} \caption{(Color online) Decoherence rates (black
circles) as a function of the scattering length. Rates are
obtained from the slope of a linear fit, at short times, of the
curves shown in Fig. 2 in unit of $2 \hbar k$. The red region
comes from theoretical predictions of the rates, for 20000 to
50000 atoms, in order to account for number fluctuations during
the measurement and for possible systematic error in the atom
number evaluation.} \label{fig1}
\end{figure}

Below $1 a_0$ we find that noise in the lattice laser starts to
significantly contribute to the decoherence, preventing a
quantitative comparison of the observation with theory. We have
however further investigated the effect of interaction on the
decoherence of the interferometer around the zero crossing. In
this region we have used a cloud, dense enough to make the effect
of the interaction visible, but sufficiently diluted in order to
completely exclude the effect of three body losses (see below) and
to prevent the condensate from collapse for small negative $a_s$
\cite{Roati}. A condensate with $a_s=a_{in} \ne 0$ is initially
prepared. Right after the beginning of Bloch oscillations the
external magnetic field is tuned to a final value in 2~ms. This
value is kept constant for 180~ms of Bloch oscillations and for
12~ms of expansion from the lattice. We have probed the
decoherence for two different densities, that were obtained by
using two different values of $a_{in}$, $3a_0$ and $1a_0$. In both
cases, the width of the interference peak reveals a minimum at
350G (see Fig.~4). Notice that this point is in agreement with the
expected position of the zero crossing ($350.4\pm0.4$~G). The
symmetric trend of the data confirms that the decoherence depends
on the magnitude and not on the sign of the scattering length.
Notice that for $a_{in}=1~a_0$ the atoms are loaded over fewer
lattice sites, resulting in larger $\rho_i$. In this case a
slightly larger width of the peak on the minimum is measured. This
confirms that interaction induced decoherence is small but still
present. The increase in the width is compatible with our expected
resolution of 0.06 $a_0$. Further experiments will be necessary to
clarify whether dipole dipole interaction between the magnetic
dipoles of the atoms starts to play a significant role when $a_s$
is tuned to such small values.

\begin{figure}
\includegraphics[width= \columnwidth
,clip]{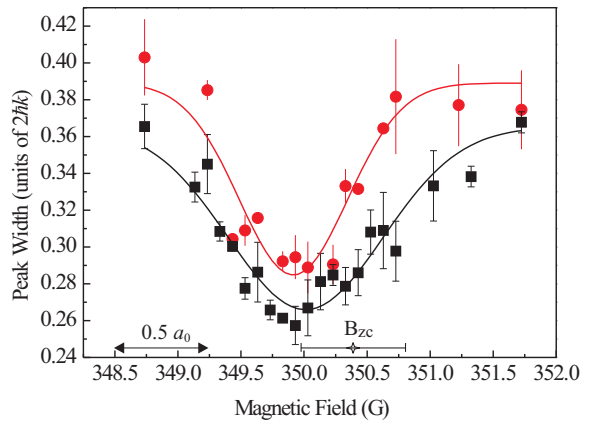} \caption{(Color online) Interference peak
width as a function of the scattering length hold after 180 ms of
Bloch oscillations. The condensate is prepared for two different
values of $a_s$, $3 a_0$ (squares) and $1 a_0$ (circles), and
tuned to the final value, in 2 ms, right after Bloch oscillations
have started. The lines represent a gaussian fit to the data. }
\label{fig1}
\end{figure}

\begin{figure}
\includegraphics[width=\columnwidth,clip]{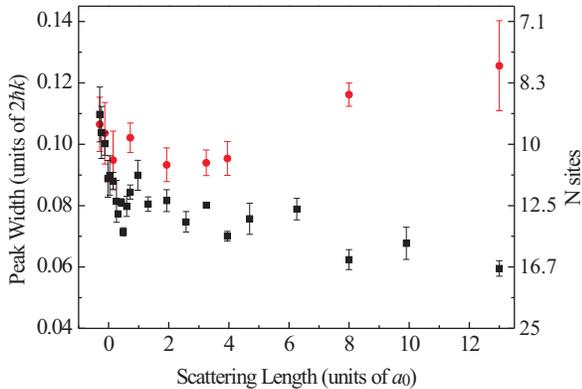}
\caption{(Color online) Width of the central peak of the atom
distribution after release from the trap and the lattice ($s=6$)
for different value of the scattering length of the condensate
during the loading of the lattice. The trapping frequencies
generated by the dipole trap and the lattice are $\nu_z = 100$~Hz,
$\nu_x = 95$~Hz, $\nu_y = 113$~Hz. The expansion time is 22 ms.
For the squares data the scattering length is tuned to zero at the
beginning of the expansion in 0.5 ms. From the inverse of the peak
width in unit of $2\hbar k$ we can estimate the number N of
lattice sites occupied. For the circles the scattering length is
left unchanged. } \label{fig1}
\end{figure}

Tuning the scattering length has another important consequence.
During Bloch oscillations, the in-trap extension of the sample
results from the spatial interference of different Wannier Stark
states $\phi_i$. Therefore the size of the cloud can have at most
a variation of the order of the extension $l$ of the single
$\phi_i$. In our case, for $s=6$ and $F=ma$, where $m$ is the mass
of the potassium atom and $a=9.8$~m/s$^{2}$, $l\sim 2 \mu$m
\cite{Meystre}. As a consequence the spatial resolution of our
interferometer depends on the initial size of the condensate if
this is larger than $l$. Tuning $a_s$ to zero allows the
condensate to occupy the ground state of the trapping potential
and by an appropriate choice of the external confinement we can
prepare very small samples. To verify the in-trap size of our
sample we have performed an experiment in which the condensate is
prepared at variable $a_s$ and then released from the combined
potential, i.e. the trap plus the vertical lattice. In fact the
coherent splitting of the condensate over several lattice sites
results in an interference pattern in the momentum space with a
central peak width in units of $2\hbar k$ approximatively equal to
the inverse of the number of occupied sites \cite{Pedri}. Notice
that the in-trap momentum distribution is accessible
experimentally, tuning $a_s$ to zero at the beginning of the
expansion. Measurements of the interference peak width as a
function of $a_s$ of the condensate are presented in Fig.5. At
$a_s=0$ the inverse of the peak width in units of 2$\hbar k$
corresponds to $\sim 10$ sites occupied. This is in agreement with
the 1/e$^2$ spatial width of 4.5 $\mu$m for a condensate confined
in our combined potential which has, for this measurement, a
vertical trapping frequency of 100 Hz. Expansions with unchanged
values of the scattering length are also reported. It is important
to stress that for small values of $a_s$ the peak density of the
cloud in presence of the lattice exceeds $10^{14}$ atoms/cm$^3$.
Three body losses \cite{Cornell} over few hundreds of ms are
negligible thanks to a low value of $K_3 = (1.3 \pm 0.5 )\cdot
10^{-29}$ cm$^6 $s$^{-1}$ for $^{39}$K, that we have measured in
the experiment in the zero crossing region. We observe that three
body recombination in the non interacting BEC results just in
losses, and not in a heating of the sample. The possibility to
create atomic clouds of micrometric sizes allows to have a spatial
resolution much better than in previous interferometers of this
kind \cite{Roati, Ferrari}. Improvement of our current sensitivity
($10^{-4} a$) can in principle lead to precise measurements of
Casimir Polder forces between atoms and surfaces and to assign new
constraints on the Yukawa type deviations from the Newtonian
gravitational laws at small distances in the 1 $\mu$m-10 $\mu$m
region \cite{Dimopulos, Ferrari}.

In conclusion we have performed atom interferometry with a nearly
non interacting BEC of $^{39}$K. Until now mean field interaction
arising from s-wave interatomic collisions was the main limitation
in trapped interferometric sensors. We proved that interaction
induced phase diffusion can be strongly suppressed by tuning the
scattering length to zero, using a broad magnetic Feshbach
resonance. The resulting sensor can have micrometric spatial
resolution and can be implemented for measurements of forces at
small distances from surfaces. This technique can be in principle
extended to other interferometric schemes and opens new exciting
perspectives in the field of trapped atom interferometry. Moreover
dynamically tuning the scattering length of the BEC could also
open new directions towards Heisenberg limited interferometry
\cite{heisemberg}.

While writing this manuscript we became aware of a similar work
performed with $^{133}$Cs atoms at the University of Innsbruck.

We acknowledge contributions by F. Marin, discussions with A.
Simoni, A. Smerzi, S. Stringari and the rest of the quantum gases
group at LENS. This work was supported by MIUR(PRIN 2006), by
EU(MEIF-CT-2004-009939), by INFN, and by Ente CRF, Firenze.

\end{document}